\documentclass[12pt]{article}
\usepackage{amssymb,latexsym}
\title{Minimal Polynomial Algorithms for  Finite Sequences}
\author{Graham H. Norton\\
Dept. Mathematics, University of Queensland\\
 Brisbane, Queensland 4072, Australia\thanks{e-mail: ghn@maths.uq.edu.au}}
 \date{15 March, 2010}
\begin{document}
\maketitle
\begin{abstract}
\noindent We show that a straightforward rewrite of  a known minimal polynomial  algorithm yields a simpler version of a recent algorithm of  A. Salagean.
\end{abstract}

\section{Introduction}
Let $K$ be a field, $n\geq 1$  and  $s=(s_1,\ldots,s_{n})$ be a finite sequence over  $K$.
The Berlekamp-Massey (BM) algorithm computes an LFSR of  shortest length $L$ and a feedback polynomial $F\in K[x]$  generating $s$, vacuously if $L=n$ \cite{Ma69}. 

We begin with the approach and basics of \cite{N95b}. 
Multiplication makes Laurent series in $x^{-1}$ into a $K[x]$-module and power series with non-zero annihilator ideal correspond to {\em linear recurring sequences}: they have a non-zero 'characteristic  polynomial' (c.p.)  \cite[Section 2]{N95b}. For finite sequences, we work with Laurent polynomials and $C\in K[x]$ 
is a {\em c.p. of} $s$ if
\begin{eqnarray}\label{cp}
C_0\cdot s_{j-d}+\cdots+C_d\cdot s_{j}= 0
\end{eqnarray}
 for $d+1\leq j\leq n$, where $d=\deg(C)\geq 0$  \cite[Definition 2.7, Proposition 2.8]{N95b}.   Any $C$ with $d\geq n$ is vacuously a c.p. of $s$. 

A c.p. $C$ of $s$ is a {\em minimal polynomial of} $s$ if $\deg(C)=\min\{\deg(D):\ D \mbox{ is a c.p. of }s\}$ and the {\em linear complexity} of $s$ is the degree of any minimal polynomial of $s$ \cite[Definition 3.1]{N95b}, \cite[Definition 2.2]{N99b}. For example, 
$D=x^{L-\deg(F)}F^\ast$ is a c.p. of $s$ --- as usual, $F^\ast$ is the reciprocal of $F$--- and $D$ is a minimal polynomial of $s$ since $D^\ast=F$ and $\deg(D)=L$.  

As far as we know, Algorithm 4.2 of  \cite{N95b} (Algorithm 3.1 below)
was the first algorithm to compute a minimal polynomial of  $s$ iteratively.
In fact, it is valid for finite sequences over a commutative unital integral domain. 

Algorithm 2.2  of \cite{ Salagean} also computes a minimal polynomial of $s$.  We show  that  these two algorithms  are closely related:  a straightforward rewrite of the former using the notation of \cite{ Salagean} yields the latter, except that we initialise a polynomial to 0 instead of 1.
Further, the rewrite uses fewer variables and is simpler. See also Remark 3.2 (iv).

We note that  \cite{N95b} and \cite{N99b} (an expository version  of \cite{N95b})  were referred to in \cite[Introduction]{NS-key}.
\section{The Inductive Construction}\label{constr}
The whole process of  Algorithm 3.1 is best explained in terms of the inductive construction of a minimal polynomial  of $s$ which was derived from first principles in \cite{N95b}.

\subsection{The Naive Version}
A natural choice for $C^{(1)}$ is 1 if $s_1=0$ and $x$ otherwise; $C^{(1)}$ is certainly a c.p. of minimal degree. Now assume inductively that $2\leq i\leq n$ and we have c.p.'s $C^{(j)}$ for $(s_1,\ldots,s_j)$ where $1\leq j\leq i-1$. From Equation (\ref{cp}), $C^{(i-1)}$  is a c.p. of $(s_1,\ldots,s_i)$ if and only if the {\em discrepancy}   
$$c_{i-1}={\mathrm d}(C^{(i-1)})= \sum_{j=0}^{d_{i-1} }C^{(i-1)}_j\cdot s_{j+i-d_{i-1}}\in K$$
is zero, where $d_{i-1}=\deg(C^{(i-1)})$ \cite[Definition 2.10]{N95b}, {\em cf.} \cite[Equation (10)]{Ma69}.
If $c_{i-1}=0$ then   clearly $d_{i-1}$ is minimal. 
But if  $c_{i-1}\neq 0$, a new c.p. is needed. We use an {\em index} $a_{i-1}$ suggested by  \cite[Equation (11)]{Ma69} 
\begin{eqnarray}\label{ante}
a=a_{i-1}=\max_{1\leq j\leq i-2}\{j:\ d_j<d_{i-1}\}
\end{eqnarray}
to index a previous c.p. 
\cite[Definition 3.12]{N95b}. The {\em exponent}  $e_{i-1}=2d_{i-1}-i$  and $C^{(i)}$ are from \cite[Proposition 4.1]{N95b}:
\begin{eqnarray}\label{maincase} C^{(i)}=\left\{ \begin{array}{ll}
c_{a}\cdot C^{(i-1)}-c_{i-1}\cdot x^{+e}C^{(a)} \mbox { if } e=e_{i-1}\geq 0\\\\
c_{a}\cdot x^{-e}C^{(i-1)}-c_{i-1}\cdot C^{(a)} \mbox { otherwise.}
\end{array}
\right.
\end{eqnarray}
Now $a=a_{i-1}$ and Equation (\ref{maincase}) require $i\geq 3$ and $d_{i-1}>d_{1}$. For $i\geq 2$ and $d_{i-1}=d_{1}$, we complete  our construction by\begin{eqnarray}\label{minorcase} C^{(i)}=\left\{ \begin{array}{ll}
x^i &\mbox{ if } s_1=0\\
s_1\cdot x^{i-2}C^{(i-1)}-c_{i-1} &\mbox { otherwise.}
\end{array}
\right.
\end{eqnarray}
The case $s_1=0$ occurs when $s$ has $i-1\geq 1$ leading zeroes. \\

Part (i) of the following proposition is an analogue of \cite[Equation (15)]{Ma69}. For Part (ii), see \cite[Propositions 4.3, 4.5]{N99b}.

 {\bf Proposition 2.1} For $2\leq i\leq n$, if $c_{i-1}\neq 0$ then (i) $d_{i}=\max\{d_{i-1},i-d_{i-1}\}$ and
(ii)  $C^{(i)}$ is a c.p. of $(s_1,\ldots,s_i)$.\\

An analogue of \cite[Lemma 1]{Ma69} establishes minimality.

{\bf Lemma 2.2} (\cite[Theorems 3.8, 3.13]{N95b} or \cite[Lemma 5.2]{N99b})  Let $2\leq i\leq n$, $f$ be a c.p. of $(s_1,\ldots,s_{i-1})$ and ${\mathrm d}(f)\neq 0$. If $g$ is a c.p. of $(s_1,\ldots,s_{i})$, then $\deg(g)\geq i-\deg(f)$.\\ 

We now have a naive inductive construction for a minimal polynomial of $s$ and illustrate it with two binary examples.\\

{\bf Example 2.3} Consider the subsequence $s=(0,1,1,0)$ of \cite[Table I]{Salagean}. We have $C^{(1)}=1$ as $s_1=0$ and  $c_1=s_2\neq 0$. As $i<3$, we apply Equation (\ref{minorcase}): $C^{(2)}=x^2$ (there is one leading zero). Now $i=3$, $c_2=s_3\neq 0$ and $d_2=2>d_1=0$, so that $a_2=1$. Equation (\ref{maincase}) applies with exponent $e_2=2\cdot 2-3=1$ giving  $C^{(3)}=C^{(2)}+x^1 C^{(1)}=x^2+x$. Finally,  $c_3=s_4+s_3\neq 0$ and $d_3=2>d_1=0$, so that $a_3=a_2=1$. Equation (\ref{maincase}) applies again with exponent $e_3=2\cdot 2-4=0$ and
$C^{(4)}=C^{(3)}+ x^0C^{(1)}=x^2+x+1$.\\

{\bf Example 2.4} Let $s=(1,1,0,0)$. Clearly $C^{(1)}=x$ and $c_1=s_2\neq 0$, so $C^{(2)}=x^0C^{(1)}+1=x+1$ from Equation (\ref{minorcase}). Next, $c_2=s_3+s_2\neq 0$ and Equation (\ref{minorcase}) obtains as $d_2=d_1=1$, giving $C^{(3)}=x^1C^{(2)}+c_2=x^2+x+1$. For $i=4$, $c_3=s_4+s_3+s_2\neq 0$, $d_3=2>d_1=1$ and $a_3=2$.  Equation (\ref{maincase}) with exponent $e_3=2\cdot2-4=0$ gives $C^{(4)}=C^{(3)}+x^0C^{(2)}=x^2$.\\

\subsection{The Refined Version}
The naive construction can be refined in three ways. Firstly, by noting that  if $e_{i-1}\geq 0$ then $a_i=a_{i-1}$ since $d_{i-1}\geq i-d_{i-1}$ i.e. 
$d_{i}=d_{i-1}$ by Proposition 2.1. But if $e_{i-1}<0$,
$d_{i}=i-d_{i-1}>d_{i-1}$ and  $a_i=i-1$. 

Secondly, we update $e_{i-1}=2d_{i-1}-i$ (and so avoid using $d_{i-1}$) as follows: it is trivial that $e_i=e_{i-1}-1$ if $c_{i-1}=0$ and easy to check that $e_i=|e_{i-1}|-1$ if $c_{i-1}\neq 0$.

Thirdly, we change the inductive basis by introducing  artificial values $C^{(0)}=1$ (which is only a c.p. if $s_1=0$), $C^{(a_0)}=0$ (which is  not a c.p. by definition) and   $c_{a_0}=1$ (which is not a discrepancy). For definiteness, we take $a_0=-1$. Then remarkably (i)  $e_0=-1$ (ii) Equation (\ref{maincase}) accomodates all three cases and (iii) the updating of $a_{i-1}$ and $e_{i-1}$  remains valid. We state this formally.

{\bf Theorem 2.5} (\cite{N95b})
Put $a_0=e_0=-1$,  $C^{(-1)}=0$, $c_{-1}=1$ and $C^{(0)}=1$. 
For $1\leq i\leq n$, define $C^{(i)}$, $a_i$, $e_i$  by

(i) if $c_{i-1}=0$, $C^{(i)}=C^{(i-1)}$, $a_i=a_{i-1}$  and $e_i=e_{i-1}-1$  

(ii)  if $c_{i-1}\neq 0$ let $C^{(i)}$ be as in Equation (\ref{maincase}), $a_i=a_{i-1}$ if $e_{i-1}\geq 0$, $a_i=i-1$ if $e_{i-1}<0$ and $e_i=|e_{i-1}|-1$.

Then for $1\leq i\leq n$,  $C^{(i)}$ is a minimal polynomial of $(s_1,\ldots,s_i)$ and $\deg(C^{(i)})=\frac{e_i+i+1}{2}$. \\

The BM algorithm decodes Reed-Solomon, Goppa and negacyclic codes \cite{Be68}, \cite{P} and has been extended to multiple sequences \cite{FT91}. For similar applications and extensions of Theorem 2.5, see  \cite[Section 8]{N99b}, \cite{N95c}.
\section{The Algorithm}
We deduce Algorithm 3.1 from Theorem 2.5. 
Firstly, we can dispense with the indices $a_{i}$ if we define $B^{(i)}=C^{(a_i)}$ and scalars
$b_i={\mathrm d}(C^{(a_i)})$ for $0\leq i\leq n-1$ provided we update $B^{(i-1)}$ and $b_{i-1}$  when $e_{i-1}<0$.  
Secondly, we replace $d_{i-1}$  by $\frac{e_{i-1}+i}{2}$ in $c_{i-1}$. Finally,  since only current values are used we can suppress all indices --- provided we keep a copy $T$ of $C^{(i-1)}$ to update $B^{(i-1)}$  when $e_{i-1}<0$.
\\

\noindent {\bf Algorithm 3.1} (\cite[Algorithm 4.2]{N95b} rewritten)

\noindent Input:  $n\geq 1$ and a sequence $s=(s_1,\ldots,s_{n})$  over a field $K$.

\noindent Output: $C$, a monic minimal polynomial  for $s$.

\noindent {\bf begin}  $B \leftarrow 0;\ b \leftarrow  1;\ C \leftarrow 1;\ e\leftarrow-1;$
\begin{tabbing}
{\bf for} \= $i = 1$ {\bf to }$n$ {\bf do}\\
    \> $c  \leftarrow \sum_{j=0}^{\frac{e+i}{2}} C_j \cdot  s_{j+
    \frac{i-e}{2}};$ \\
   \> {\bf if} $c  \neq  0$ {\bf then  if} $e\geq 0$  \=   						{\bf then} $C \leftarrow b\cdot C-{c} \cdot x^e B;$\\  \>                 \>    {\bf else} \=$T\leftarrow C;\ e\leftarrow-e;$\\
  \>                 \>           \> $C\leftarrow b\cdot x^eC-{c} \cdot B$;\\
  \>                 \>            \> $B\leftarrow T$; $b\leftarrow c$;\ {\bf endif}\\
  \>                 \> {\bf endif}\\
  \> $e\leftarrow e-1$;\\
 {\bf endfor} \\   
{\bf return} ($C/b)$ {\ \bf end.}  \ \
\end{tabbing}

{\bf Remark 3.2} We obtain \cite[Algorithm 2.2]{ Salagean} except that \\

(i) we could (but do not) make each $C$ monic 

(ii) in the notation of \cite[p. 4696]{Salagean}, we  do not keep track of  '$m$' or $\deg(B)$ to recompute  $v=N-m-(\deg(C)-\deg(B))$  at each iteration (where $0\leq N\leq n-1$)

(iii)  in fact, (a) $N$ corresponds to $i-1$ (b) if there are $N$ leading zeroes $v=N+1$ (c) if $\deg(C)>\deg(B)$ then $\deg(C)=m+1-\deg(B)$ and
$ v=N-m-(\deg(C)-\deg(B))=N+1-2\deg(C)$; thus $v$  corresponds to $-e$

(iv)  $B^{(0)}=0$ and some initial minimal polynomials will   differ;  with $B^{(0)}=C^{(-1)}=1$, $C^{(i)}=x^i-c_{i-1}$ when there are $i-1\geq 0$ leading zeroes, Theorem 2.5 remains valid, the algorithms are equivalent and their outputs are identical. \\

{\bf Remarks 3.3 (added in proof)}\\

(i) Theorem 2.4 of  \cite{Salagean} on the set of minimal polynomials of $s$ is an immediate consequence of 
\cite[Theorem 4.16]{N95b}: simply replace $d$ in Theorem 4.16, {\em loc. cit.} by $-v=\deg(C)-\deg(B)-(n-m)$  and take first components.

(ii) In \cite{N2010a}, we calculate the monic {reciprocals} of the minimal polynomials of Theorem 2.5. This readily yields an algorithm similar to the BM algorithm, except that we do not calculate the 'lengths' $L_i$; we update the exponents as above and obtain $L_n$ as $\frac{e_n+n+1}{2}$. This article also contains proofs for the results of Section \ref{constr} using similar notation.\\

The author thanks the anonymous reviewers for their helpful comments and suggestions.

\end{document}